\documentclass{nature}
\usepackage{textcomp}
\usepackage{gensymb}
\usepackage{amsmath}
\usepackage{multimedia}
\usepackage{graphicx}
\usepackage{placeins}
\usepackage{color,soul}

\usepackage{color} 
\usepackage{footnote}
\makesavenoteenv{tabular}
\makesavenoteenv{table}
\usepackage{amsfonts}

\usepackage{graphicx}
\makeatletter
\let\saved@includegraphics\includegraphics
\AtBeginDocument{\let\includegraphics\saved@includegraphics}
\renewenvironment*{figure}{\@float{figure}}{\end@float}
\makeatother

\title{Quantification of collective behaviour via causality analysis}

\author{Kirill Lonhus, Dalibor \v{S}tys \& Renata Rycht\'{a}rikov\'{a}}

\begin{document}
\maketitle

\begin{affiliations}
 \item University of South Bohemia in \v{C}esk\'{e} Bud\v{e}jovice, Faculty of Fisheries and Protection of Waters, South Bohemian Research Center of Aquaculture and Biodiversity of Hydrocenoses, Institute of Complex Systems, Z\'{a}mek 136, 373 33 Nov\'{e} Hrady, Czech Republic.
\end{affiliations}

\begin{abstract}
Terms such as \textit{leader}, \textit{mediator}, and \textit{follower} sound equal in the description of a pack of wolves, a street protest crowd, or a business team and have very similar meanings. This indicates the presence of some general law or structure that governs collective behaviour. To reveal this, we selected the most common parameter for most levels of the organisation -- motion. A causality analysis of distance correlations was performed to obtain follow-up networks that show who follows whom and how strongly. These networks characterise an observed system in general and work as an automation bridge between the biological experiment and the broad field of network analysis. The proposed method was tested on 3D image data from a controlled experiment on a 6-member school of aquarium fish of Tiger Barb. The network patterns can be easily ethologically interpreted and agreed with expected behaviour.
\end{abstract}



\section{INTRODUCTION}




Collective behaviour is one of the most enigmatic phenomena of contemporary times \cite{BakColeman2021}. It is a genuinely omnipresent phenomenon -- from the flocking of birds and the growth of bacterial colonies \cite{Gibbs2008,Dinet} to the stock market trading and hierarchy of school classes. We intuitively feel that something should be familiar to all these phenomena because the observed patterns of agents' behaviour, such as \textit{leader}, \textit{follower}, \textit{mediator}, or \textit{crowd}, sound applicable to all these systems. There is probably some general law or structure common to most living species and unrelated to the degree of evolution and the level of the organisation \cite{Vicsek2012, Vicsek2}.

This work justifies and formalises such a general meaning and empirically verifies the findings. To be able to link such diverse types of collective behaviour, it is necessary to choose its common manifestation. We selected \textit{the motion}. This term has a similar meaning for most groups of living species (a move from "bad" to "good", whatever this means) and is invariant to the organisation level (degree of evolution). For instance, cells move from a place without nutrients to a place with nutrients or dogs run away from vehicles.

One of the most common approaches to evaluating motion is the analysis of agents' relative position \cite{Brown2018a}. This method was successfully applied in the investigation of human behaviour \cite{Warren2018} and fish \cite{RomeroFerrero2019}. There are two types of this approach: physically inspired and model-free. The first one assumes a direct analogy with the physical system: agents are considered particles that are driven by potential -- spring-like -- forces. These potential-driven particles are the first models that can explain the behaviour of flocks and schools \cite{Vicsek2012}. This set of techniques is an excellent tool for quantifying known phenomena but not discovering new ones.

Another -- data-driven -- approach characterises the agents' behaviour by specific, measurable quantities (e.g., the nearest neighbour distance, the Voronoi cell size) and then uses these high-level features to describe observed phenomena. To a certain extent, this approach works, but a more recent paper \cite{StrandburgPeshkin2013} proved it insufficient. However, later, the data-driven method using the diffusion maps was applied successfully to identify states of collective behaviour in dense fish schools~\cite{Titus}.

The next generation of models \cite{Davidson2021a} defined a network based on each agent's position and the sensory field. This step was necessary for incorporating the behaviour itself (the individual's reaction to the observed situation) into the model that would address the decision-making process. However, this approach is still not fully model-free. It assumes that the sensory field is tied to the agent, with the interactions being semi-local and homogeneous. Additionally, these interactions are predefined by, e.g., a similar orientation or moving in the same direction. Despite the remarkable success, extending this approach to include heterogeneous agents and non-local interactions is challenging. 

To address these limitations, we propose a more model-free way of characterising a group of interacting agents. This approach does not limit interactions by type or locality. Moreover, as we will show in the results, this approach allows the agents to be heterogeneous, which is crucial to explaining the biological experiment results. Due to causality analysis, we are not enforcing the type of pattern (e.g. moving in the same direction), only rotation and scale invariant pattern repeatability.

We selected a small aquarium fish -- Tiger Barb -- as a testing subject. This species is very active and smart, able to form shoals from 5 individuals with complex collective behaviour \cite{Saxby}. We employed video analysis to track each fish's movement without any label.

Here we propose a method of characterising collective behaviour by causality networks. It is a way of converting time-resolved multi-agent dynamics to a simple mathematical graph, showing who follows who. Such a follow-up network can be calculated for a time window, outlining the system's time evolution. This approach is generic, considering only Euclidean coordinates of the objects. However, this approach requires knowledge of the trajectories of each agent. Thus, tracking and recognition of individuals are needed.

\section{Materials and methods}

\subsection{Fish breeding}\mbox{}\\
As a model organism, we used a Tiger Barb (\textit{Puntigrus tetrazona}) in a group of 6 individuals.

Between the experiments, the fish lived in a relatively large home aquarium (60$\times$35$\times$40 cm$^3$). The aquarium was disinfected with ethanol, boiling, and sodium hydrogen carbonate and designed to simulate the natural environment with involved aquatic plants and stones (4--8 mm). The fish were bred at a 12/12 artificial daylight cycle and 25 $\degree$C. The water quality was validated at weekly intervals and before each experiment. The parameters of water were temporarily adjusted according to the conditions prevailing in the aquarium: 4.9 g\,$\cdot$\,L$^{-1}$ Mg$^{2+}$, 12.0 g\,$\cdot$\,L$^{-1}$ Ca$^{2+}$, 5.5 g\,$\cdot$\,L$^{-1}$ NO$^{3-}$, pH 6.9, hardness 2.8 $\degree$H. The aquarium was equipped with a Sera LED 360, a Sera Air 110 air pump, a Sera Fil Bioactive 130 L separate filter (all from Heinsberg, Germany), and a water heater. The fish were fed (predominantly by the Tetra Min, Tetra, Melle, Germany) once a day (at 7 a.m.), always 1 h after turning on the aquarium illumination.

\subsection{Image data collection}
\begin{figure}[t]
\centering
\includegraphics[width=\linewidth]{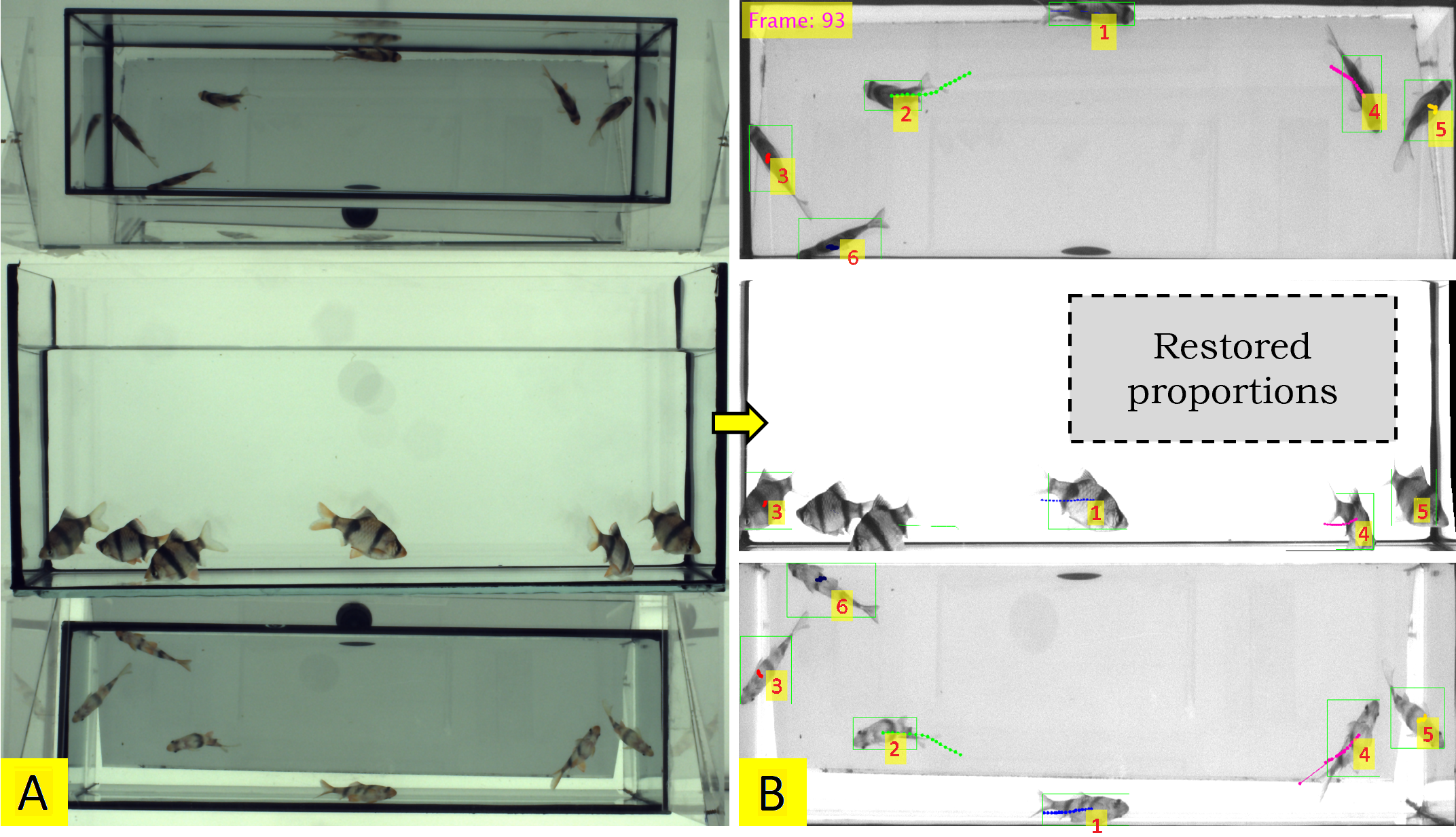}
\caption{Aquarium.
\textbf{(A)} The experimental setup (a fish aquarium with two side mirrors) and \textbf{(B)} individual objects (fish) tracking in the multiple views after image geometry restoration.}
\label{fig1}
\end{figure}

\mbox{}\\
The experiments were conducted at (11 $\pm$ 0.5) a.m. on 10 consecutive days (including the weekend). The condition of the fish was assessed before each experiment. Figure~\ref{fig1} represents the experimental set-up. The small ($38.5\times20\times14$ cm$^3$; maximal volume 8-L; 6-L used) aquarium was located inside a LED box, which provided homogeneous illumination from all angles. Water in the aquarium came from the home aquarium. Along four sides of the aquarium, mirrors were located. Two -- above and below the aquarium -- mirrors were sufficient to observe the aquarium from 3 sides simultaneously (one side directly and two sides in mirrors). The fish were transported to the small aquarium under full illumination 5 min before each experiment. Of course, such a setup may cause significant stress to the fish. However, the previous exposure of the fish to light significantly mitigated the behaviour alteration. In addition, this species has a keen intellect and good memory, so the experimental conditions are not new for it every day, and the stress is alleviated. Above all, the aim was not to fully eliminate the stress but to make it repeatable and systematic so that all fish perceive it equally. All experiments were performed with a single group of fish to reduce inter-group variability. It is unknown how different school structures can be (even for the same species and group size).

To reduce the errors in identifying individual fish, the data was collected by a 12-bit Ximea MX124CG-SU-X2G2-FAB rgb camera (scanning frequency 0.4 Hz). The size of the fish in the digital image corresponded to approx. 50$\times$30 px (5 \textmu m$^2\,\cdot$\,px$^{-1}$).

\section{Results}
\subsection{Fish tracking}\mbox{}\\
During the experiment, the aquarium with fish was recorded with a frequency of 25 Hz. The images were stored uncompressed in the 16-bit depth. Before further processing, we had to separate the camera images into corresponding views. The aquarium corners were annotated manually (a total of 28 points) and then fitted with a projective transformation. As a side effect, the projective transformation compensated for the distortion of the corresponding views (Figure~\ref{fig1}) and the aspect ratio became linearly proportional to the actual size of the aquarium.

\begin{figure}[t]
\centering
\includegraphics[width=\linewidth]{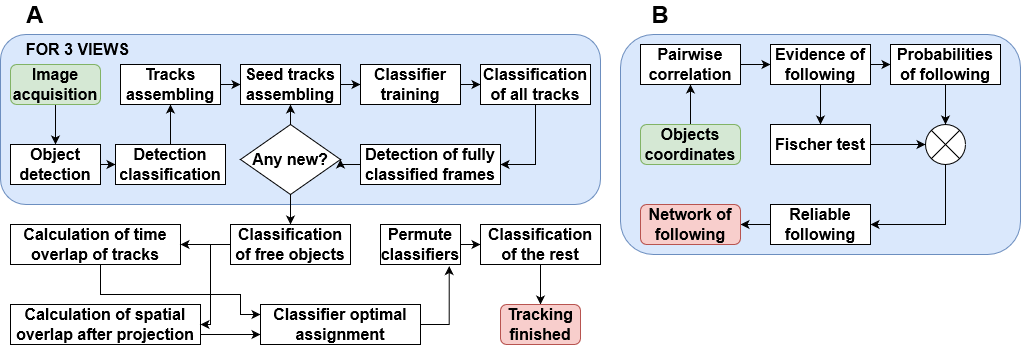}
\caption{Algorithm.
\textbf{(A)} The extraction of objects' coordinates and tracking algorithm. \textbf{(B)} The algorithm for construction of follow-up networks. This algorithm arises from the tracking algorithm \textbf{A}.}
\label{fig2}
\end{figure}

The aquarium front view giving the $z$-coordinates of the fish localization was always processed separately. The bottom and top views were almost the same but only mutually flipped (in terms of coordinates $x$, $y$) and, thus, were usually averaged. If the fish were overlapping in the top view and not in the bottom view (or vice versa), the bottom and the top view were processed separately to give better tracks.
The general method pipeline (Figure~\ref{fig2}A) was inspired by Perez-Escudero et al. \cite{Perez-Escudero2014} and~Romero-Ferrero et al. \cite{RomeroFerrero2019} but rewritten from scratch. The original algorithm is 2D, but, in our case, three 2D views allow us the reconstruction of 3D positions (with $x$-$y$ redundancy when one view is from above and the second one is from below the aquarium) using re-projecting the coordinates and calculating the overlap of coordinates between the assembled trajectories in different views. This allows us to permute the classifiers that were the first (for all views) to detect the same fish in the corresponding views. In this way, we get a consistent annotation between views.

The first step used a simple foreground detection based on median background estimation. This foreground detection is very robust and shows a high recall. The detected objects were then classified as individuals/overlaps using a small VGG-like CNN \cite{Liu2015}. This CNN annotation was generated automatically based on the average areas and perimeters of the detected objects. The detections for each fish were assembled into contiguous sequences -- tracklets -- based on the overlapping bounding boxes in every two consecutive frames. In some cases, it was possible to resolve the overlaps of two fish \cite{Lonhus2019} to increase the continuity of the tracklets.

The image sequences, where the number of detected objects coincides with the number of fish in the experiment, were used as seeds for the next CNN-based classifier. These sequences for the different aquarium sides were matched using the projection of coordinates and overlap in the time domain \cite{Perez-Escudero2014} and used to train the classifier. Then, all objects were classified. Among them, the trustworthy sequences were selected and used for training. Four passes were typically enough to achieve a high (0.98) classification accuracy. In the same way, the whole tracklets were classified. The tracklets from different views were projected (using known proportion between coordinates) and merged, giving 3D coordinates [mm].

Specific post-processing steps were applied to fill the gaps in the trajectories (where it was possible to do it unambiguously via overlap uniqueness) by a simple linear interpolation between the ends of the individual tracklets. We verified the method's accuracy only visually. However, the method was designed similarly to~Perez-Escudero et al. \cite{Perez-Escudero2014} and, thus, its accuracy is assumed to be comparable. Due to the design of the method, visual inspection of the fish trajectories has not shown any significant errors. The output of this step is a time-resolved 3D trajectory of each fish. We intentionally did not describe the tracking algorithm in-depth because the processing and analysis below are not narrowly limited to this data but are also suitable for any position tracking data. The complete code with comments is available in supplementary materials.

\subsection{Determination of temporal correlations}\mbox{}\\
To compare the results for different species/levels of the organization, we needed to decouple the behaviour from the motion. A classical way to do so is to use a scalar product of the direction vector as a measure of movement correlation~\cite{Nagy2010}. However, this approach covers only orientation but no speed or overall travelling distance. A more holistic way is to use a distance correlation, which also considers a non-trivial translation and handles the complex motion much better. The idea was simple: rewrite the classical covariance by multiplying the signed distances. Let $A, B$ be scalar sequences of length \textsf{N}, then
\begin{equation} \label{cov}
\mbox{cov}(A, B) = \frac{1}{\textsf{N}^2}\sum_{\textsf{i}}^{\textsf{N}} \sum_{j}^{\textsf{N}} (A_\textsf{i} - A_\textsf{j})\cdot (B_\textsf{i} - B_\textsf{j}) = \frac{1}{\textsf{N}^2}\sum_{\textsf{i}}^{\textsf{N}} \sum_{\textsf{j}}^{\textsf{N}} D_{\textsf{ij}}(A)\cdot D_{\textsf{ij}}(B)
\end{equation}
where $D_\textsf{{ij}}(X)$ is a signed distance between elements \textsf{i} and \textsf{j} of arbitrary sequence $X$. If we replace such a signed distance by the Euclidean distance $D_\textsf{ij}(\vec{X}) = ||\vec{X}_\textsf{i} - \vec{X}_\textsf{j}||$, the obtained function will behave very similar to the classical covariance and let us denote it as a distance covariance. Such a definition allows us to introduce a Pearson-like correlation coefficient, which we denoted the distance correlation, by replacing the covariances with the distance covariances:
\begin{equation} \label{corr}
\mbox{corr}(\vec{A}, \vec{B})= \textsf{N}^2 \frac{\sum_{\textsf{i}, \textsf{j}} \lVert \vec{A_\textsf{i}}-\vec{A_\textsf{j}}\rVert\cdot \lVert \vec{B_\textsf{i}}-\vec{B_\textsf{j}}\rVert}{
\sqrt{(\sum_{\textsf{i},\textsf{j}} \lVert \vec{A_\textsf{i}}-\vec{A_\textsf{j}}\rVert^2)\cdot (\sum_{\textsf{i},\textsf{j}} \lVert \vec{B_\textsf{i}}-\vec{B_\textsf{j}}\rVert^2)}},
\end{equation}
where $\vec{A}, \vec{B}$ are vector sequences with the length of \textsf{N} = \{..., \textsf{i}, \textsf{j},...\} points in the Euclidean space of the same dimensionality. Despite its simplicity, this method was rigorously done and investigated quite recently \cite{Szekely2007}. However, it varies from $[0, 1]$ instead of $[-1, 1]$, which is a direct consequence of the unsigned distance. As shown in an even more recent paper \cite{Szekely2009}, such characteristics show most of the properties and common meaning of the ordinary correlation.

\begin{figure}[t]
\centering
\includegraphics[width=\linewidth]{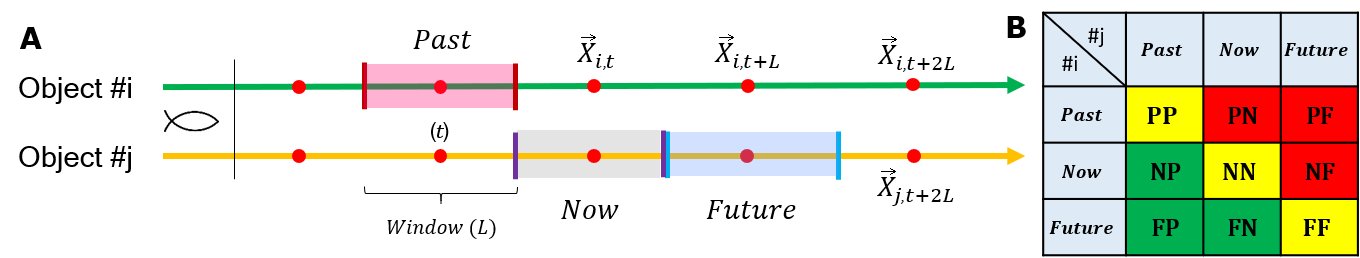}
\caption{Time windows.
\textbf{(A)} The timelines for two objects with relevant correlation time windows of the length $L$. \textbf{(B)} The favourable (\textbf{green}), neutral (\textbf{yellow}), and unfavourable (\textbf{red}) combinations of correlations for a situation when object \textit{i} follows object \textit{j}. Symbols \textbf{P}, \textbf{N}, and \textbf{F} denote the past, current, and future states of the object $i$ or $j$, respectively.}
\label{fig3}
\end{figure}

We used the correlation in Equation~(\ref{corr}) as a measure of the trajectories' similarity in the most general sense. The correlation between positions in a sliding window for two selected objects on a timeline gave the local similarity of their motion (Figure~\ref{fig3}A).

It is also possible to select windows mutually shifted on the time arrow. For an object, let us select three dedicated time windows centred at times $t$, $t-L$, and $t+L$, respectively, where $L$ is a window size, and denote them as \textbf{P}\textit{ast}, \textbf{N}\textit{ow}, and \textbf{F}\textit{uture}. Thus, any pair of objects gives 9 combinations for two windows. Let us denote the time windows by two-letter codes: the first letter marks the first object's time state, and the second letter is for the second object's time state. For example, $\textbf{PF}$ means a comparison of the first object's past position with the second object's future position. This notation helped us to analyse the physical interpretation and causality of such shifted correlations.


\subsection{Construction of follow-up networks}\mbox{}\\
Let us define the term \textit{following} in this way: if object \textit{i's future state} correlates with object \textit{j's past state}, then \textit{object i follows object j}.

If we consider the hypothesis that $i$ follows $j$ then combinations of time windows can be classified into 3 groups (Figure~\ref{fig3}B): supporting the hypothesis $S=\{$\textbf{NP}, \textbf{FP}, \textbf{FN}$\}$, neutral to the hypothesis $N=\{$\textbf{PP}, \textbf{NN}, \textbf{FF}$\}$, and rejecting the hypothesis $R=\{$\textbf{PN}, \textbf{PF}, \textbf{NF}$\}$. If each time window is taken from each group only once, it gives $3^3$ combinations. We denoted as an evidence the fact that one correlation is higher than another. Such an evidence is treated as significant if the corresponding correlation is higher than the correlation from the neutral group. In the next step, we evaluated the statistical significance of the hypothesis whether the tendency that object $i$ follows object $j$ and vice versa is higher than that object $j$ follows object $i$. For each pair of objects $i$ and $j$, we have four possibilities how the objects follow each other in a time window: $i$ follows $j$; $j$ follows $i$; $i$ does not follow $j$; $j$ does not follow $i$. The algorithm for constructing the follow-up networks is depicted in Figure~\ref{fig2}B.

The total numbers of the significant evidences during the experiment are

\begin{equation}
e^{(i \rightarrow j)} = \sum_{s^{(i \rightarrow j)}}^{S^{(i \rightarrow j)}} \sum_{n}^{N} \sum_{r^{(i \rightarrow j)}}^{R^{(i \rightarrow j)}} [s^{(i \rightarrow j)} > n]\cdot [s^{(i \rightarrow j)} > r^{(i \rightarrow j)}],
\label{Eq3}
\end{equation}

\begin{equation}
e^{(j \rightarrow i)} = \sum_{s^{(j \rightarrow i)}}^{S^{(j \rightarrow i)}} \sum_{n}^{N} \sum_{r^{(j \rightarrow i)}}^{R^{(j \rightarrow i)}} [s^{(j \rightarrow i)} > n]\cdot [s^{(j \rightarrow i)} > r^{(j \rightarrow i)}],
\label{Eq4}
\end{equation}

\begin{equation}
e^{(i \not\rightarrow j)} = \sum_{r^{(i \not\rightarrow j)}}^{R^{(i \not\rightarrow j)}} \sum_{n}^{N} \sum_{s^{(i \not\rightarrow j)}}^{S^{(i \not\rightarrow j)}}  [r^{(i \not\rightarrow j)} > n]\cdot [r^{(i \not\rightarrow j)} > s^{(i \not\rightarrow j)}],
\label{Eq5}
\end{equation}

\begin{equation}
e^{(j \not\rightarrow i)} = \sum_{r^{(j \not\rightarrow i)}}^{R^{(j \not\rightarrow i)}} \sum_{n}^{N} \sum_{s^{(j \not\rightarrow i)}}^{S^{(j \not\rightarrow i)}}  [r^{(j \not\rightarrow i)} > n]\cdot [r^{(j \not\rightarrow i)} > s^{(j \not\rightarrow i)}],
\label{Eq6}
\end{equation}
where $e^{(i \rightarrow j)}$, $e^{(j \rightarrow i)}$, $e^{(i \not\rightarrow j)}$, and $e^{(j \not\rightarrow i)}$ are sums of significant evidences when object $i$ follows object $j$, object $j$ follows object $i$, object $i$ does not follow object $j$, and object $j$ does not follow object $i$, respectively. Variables $s^{(i \rightarrow j)}$ and $s^{(j \rightarrow i)}$ denote correlations supporting the hypothesis that object $i$ follows object $j$ and vice versa, respectively. The analogous variables $r^{(i \rightarrow j)}$ and $r^{(j \rightarrow i)}$ denote correlations rejecting the hypothesis that object $i$ follows object $j$ and vice versa, respectively. Variable $n \in N$ is a combination of time windows that is neutral to the hypothesis of following up.

The discrete sums of the significant evidences in Equations~(\ref{Eq3})--(\ref{Eq6}) were first used as an input into the Fisher exact test \cite{Fisher1922} to calculate the precise $p$-value. In this case, the meaning of the $p$-value is a measure of the deviation from the hypothesis that no one follows anyone. From Table~\ref{Tab1} the $p$-value of the Fisher test was calculated as

\begin{eqnarray}
p &=& \frac{\binom{e^{(i \rightarrow j)} + e^{(j \rightarrow i)}}{e^{(i \rightarrow j)}}\binom{e^{(i \not\rightarrow j)} + e^{(j \not\rightarrow i)}}{e^{(i \not\rightarrow j)}}}{\binom{e^{(i \rightarrow j)} + e^{(j \rightarrow i)}e^{(i \not\rightarrow j)} + e^{(j \not\rightarrow i)}}{e^{(i \rightarrow j)} + e^{(i \not\rightarrow j)}}} \equiv \frac{\binom{e^{(i \rightarrow j)} + e^{(j \rightarrow i)}}{e^{(j \rightarrow i)}}\binom{e^{(i \not\rightarrow j)} + e^{(j \not\rightarrow i)}}{e^{(j \not\rightarrow i)}}}{\binom{e^{(i \rightarrow j)} + e^{(j \rightarrow i)}e^{(i \not\rightarrow j)} + e^{(j \not\rightarrow i)}}{e^{(j \rightarrow i)} + e^{(j \not\rightarrow i)}}} \equiv \nonumber \\ \\
&\equiv& \frac{(e^{(i \rightarrow j)} + e^{(j \rightarrow i)})!(e^{(i \not\rightarrow j)} + e^{(j \not\rightarrow i)})!(e^{(i \rightarrow j)} + e^{(i \not\rightarrow j)})!(e^{(j \rightarrow i)} + e^{(j \not\rightarrow i)})!}{e^{(i \rightarrow j)}!e^{(j \rightarrow i)}!e^{(i \not\rightarrow j)}!e^{(j \not\rightarrow i)}!(e^{(i \rightarrow j)} + e^{(j \rightarrow i)} + e^{(i \not\rightarrow j)} + e^{(j \not\rightarrow i)})!}. \nonumber
\label{eq:fisher2}
\end{eqnarray}

\begin{table}
\caption{\label{Tab1}The Fisher test contingency table for verification of the hypothesis of measure of following up.}
\begin{center}
\begin{tabular}{l | l l | l}
\hline
 & Object $i$ & Object $j$ & Total rows\\ \hline
Following & $e^{(i \rightarrow j)}$ & $e^{(j \rightarrow i)}$ & $e^{(i \rightarrow j)} + e^{(j \rightarrow i)}$\\
Not following & $e^{(i \not\rightarrow j)}$ & $e^{(j \not\rightarrow i)}$ & $e^{(i \not\rightarrow j)} + e^{(j \not\rightarrow i)}$\\
Total columns & $e^{(i \rightarrow j)} + e^{(i \not\rightarrow j)}$ & $e^{(j \rightarrow i)} + e^{(j \not\rightarrow i)}$ & $e^{(i \rightarrow j)} + e^{(i \not\rightarrow j)} + e^{(j \rightarrow i)} + e^{(j \not\rightarrow i)}$\\ \hline
\end{tabular}
\end{center}
\end{table}

The relations $i$-$j$, i.e., the time windows, whose $p$-values lie under the criterion of level of significance $\alpha=0.05$ (i.e., $p\leq C_{\alpha}$) were rejected. Removing these spurious correlations \cite{Farine2017} greatly improved the robustness of the obtained networks.

For the rest of relations $i$-$j$, the \textit{follow-up probabilities} were then defined using the sums of significant evidences written in Eqs.~\ref{Eq3}--\ref{Eq6}, e.g.,
\begin{equation}
P^{(i \rightarrow j)} = \frac{e^{(i \rightarrow j)}}{e^{(i \rightarrow j)} + e^{(i \not\rightarrow j)}} = [1+\frac{e^{(i \not\rightarrow j)}}{e^{(i \rightarrow j)}}]^{-1},
\label{eq:followij}
\end{equation}
where $P^{(i \rightarrow j)}$ is a chance that object $i$ follows object $j$. The advantages of the proposed measure of following up are its clear meaning and decoupling from direct measurements by transformation through the correlation space to the probability space. It allows us to compare and analyse very different systems because absolute correlation values do not matter, only their combinations and ratios. The obtained follow-up matrix is of the size $n_{obj}\times n_{obj}$ in the case of $n_{obj}$ objects and is typically sparse. It is natural to convert it to a directed graph (treating the following matrix as an adjacency matrix). The orientation of the network edges follows the comparison of the values of probabilities $P^{(i \rightarrow j)}$ and $P^{(j \rightarrow i)}$. Examples of such graphs (visualised in a way that a higher follow-up probability corresponds to a shorter edge) are shown in Figure~\ref{fig4}A. These graphs are defined at each recorded time (Video~in Supplementary) and have an apparent behaviouristic meaning, showing who follows whom.

\begin{figure}[t]
\centering
\includegraphics[width=\linewidth]{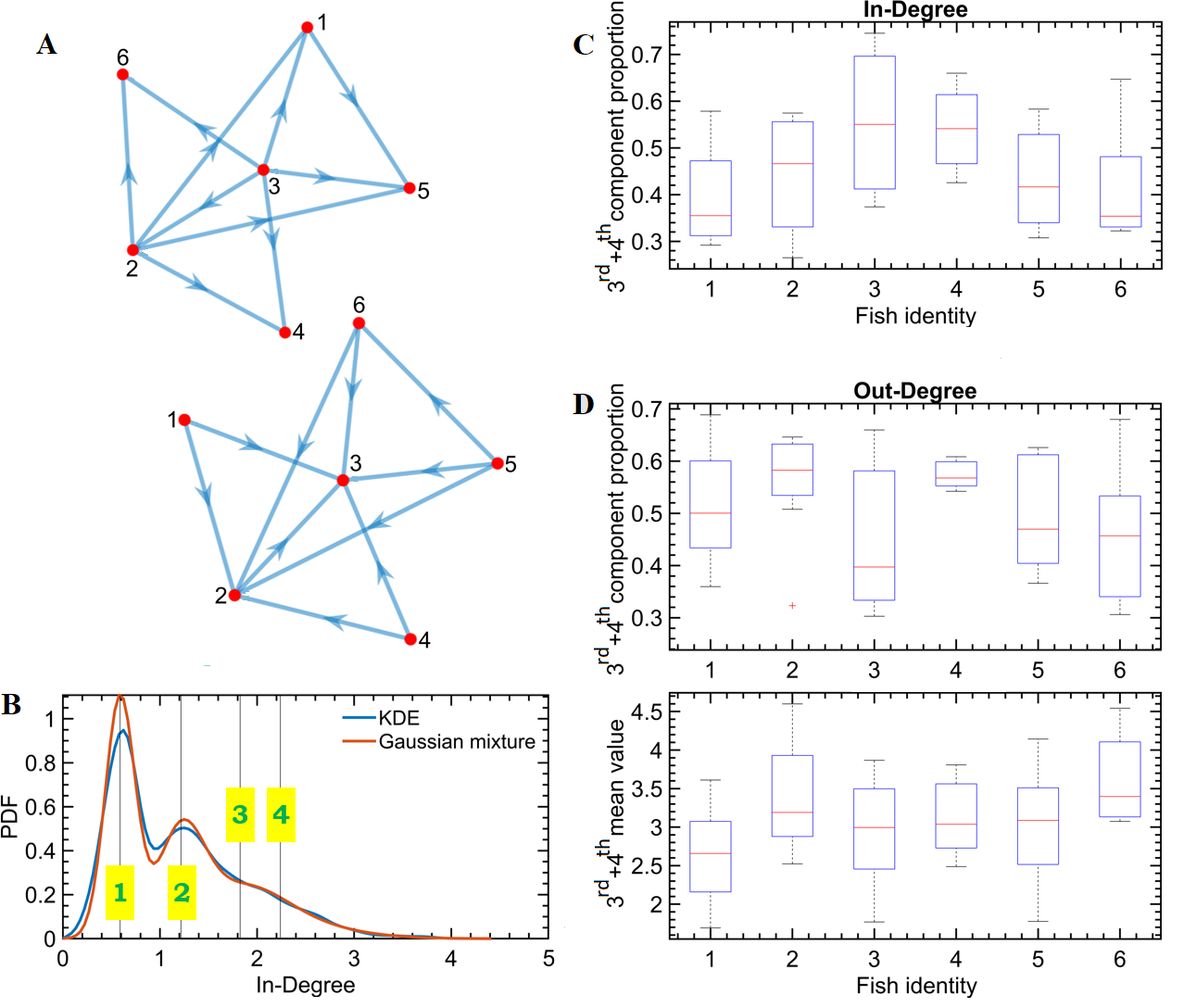}
\caption{Follow-up networks.
\textbf{(A)} Two typical follow-up networks in a school of 6 fish. \textbf{(B)} The kernel density estimation (KDE) and the Gaussian fitting for the in-degree parameter of the follow-up network. For the out-degree parameter, the distributions are analogous to the in-degree parameter. (\textbf{C}) The distribution of intra-experiment in-degree Gaussian fitting component proportions. (\textbf{D}) The distribution of intra-experiment out-degree Gaussian fitting component proportions and mean values.}
\label{fig4}
\end{figure}

The approach described above utilises only one parameter – the time window size $L$. The obtained network is thus a function only of the time $t$ and the parameter $L$.

\subsection{Biological relevance}\mbox{}\\
To verify the method, we conducted a straightforward experiment: the fish were freely swimming in an aquarium for 15 min. The experiment was repeated ten times on different days to investigate the fish school's stability and the measurements' repeatability. An output of the method is a directed graph defined for each time frame (Video~in Additional files). Interpretation of these graphs is straightforward: the nodes correspond to the observed objects (fish individuals in our case), length of the edges inversely reflects the follow-up probability (the shorter the edge, the higher the chance of following up). The arrows show who follows whom. 

To show the variability of the follow-up network over time, we partition the graphs adjacency matrices into two groups by $k$-means. Figure~\ref{fig4}A shows the follow-up networks dominating each group. Of course, the actual situation is more complicated, and two groups of follow-up networks are only the minimal illustrative example. The networks (Figure~\ref{fig4}A) are almost the same, but the directions of edges are opposite, and fish $\#1$ and $\#6$ swapped their positions. Fish $\#3$ situated in the center has the most prominent set of features among all. Even without biological analysis, one can assess that this fish is somehow important. Fish $\#3$ has relatively strong connectivity to all other fish and is completely changing its role in the school (depicted by a reverse direction of the network edges). Fish $\#2$ is also connected to everyone, but the connections are weak (the lengths of network edges are relatively long). Fish $\#5$ and $\#6$ changed their mutual distances in the graphs and did not show distinct features.

The number of edges entering/emerging from a network node in a time window characterises the in/out-degree. For each node, two centrality measures -- in-degree and out-degree -- quantify the intensity of the relations among the fish. After that, for all ten experiments, a probability density function for values of in/out-degree parameters was estimated by the kernel density estimation (Figure~\ref{fig4}B). The obtained distribution is not unimodal nor trivial. To switch to the parametric statistics, a mixture of 4 Gaussians was fitted to this distribution (Figure~\ref{fig4}B). Then, both in- and out-degree parameters can be described by four sets of [component proportion, mean value, variance] corresponding to one peak in the distribution.

The component (peak area) proportions in the probability density functions (compared with mean values or variances) provide the most straightforward interpretation of the results. This approach is closely related to frequency analysis. A higher value of the component proportion corresponds to a longer time spent in the relevant mean value (the components are sorted in ascending order according to the mean values). A robust measure of in/out-degree due to a relatively low frequency of occurrence is a sum of the 3$^{\mbox{rd}}$ and the 4$^{\mbox{th}}$ component proportion, showing how frequently the fish exhibits a high in/out-degree state. The box plot of in-degree states for all six fish from 10 experiments is depicted in Figure~\ref{fig4}C. Fish $\#3$ and $\#4$ show the highest in-degree state. These individuals are frequently followed and \textit{lead} the school. However, $\#3$ has a low out-degree and rarely follows someone, whereas $\#4$ has a high out-degree. We interpret this as meaning that fish $\#3$ is a leader, whereas fish $\#4$ can be called a coordinator. As supported in Figure~\ref{fig4}A, the issue is that there is no clear definition of possible roles, only a consensus that they exist \cite{Marras2013}. The role of the \textit{coordinator} changes diametrically (the coordinator can be followed or following). In general, such a network shows an oscillatory process over time and is not trivial (it is impossible to obtain from one cluster another by simply changing the direction of edges). We have not investigated this phenomenon deeper yet, but a similar oscillation process has already been observed before \cite{Swain2015}. The mean value, which corresponds to the out-degree Gaussian components 3 and 4 for fish $\#1$, is significantly lower than the average (Figure~\ref{fig4}D). The combination with a very low in-degree component proportion may support the hypothesis that fish $\#1$ is \textit{oppressed} or, more precisely, \textit{unsocial}. Fish $\#1$ is never a leader but also does not follow the school too much.

The typical results reflecting the mean “intensity” of the following-up (Fig.~\ref{fig4}A) may not correlate with the time statistics (frequency) of what fraction of time the agent exhibits a given value of the following-up (Fig.~\ref{fig4}C--D). Fig.~\ref{fig4}A shows that fish $\#3$ has strong connectivity as both a follower and a leader to most individuals in the school. But according to Fig.~\ref{fig4}C--D, fish $\#3$ is much more often followed than following, which supports assumptions about its role.

Thus, even this most simple analysis of the obtained graphs gives an insight into the school structure, which is consistent between the experiments. Of course, the detected \textit{roles} must be rigorously tested in different schools, even with other species. Nevertheless, the complexity of the observed phenomena is quantifiable. Our results agree with the trend recently published in research papers \cite{MarDelgado2018,Marras2013,Shaw2020} that variability in individual behaviour matters in collective behaviour. We think that this variability is essential to explain the observed phenomena. But further studies are still needed, including a detailed comparison of the obtained results with visual observations.

\section{Discussion and conclusions}
Commercially available deep-learning-based devices~\cite{Noldus} offer to track and analyse only one animal individual in 2D. Indeed, the number of research papers describing fish schools reaches hundreds yearly but none of the experiments is followed by the automatic network analysis. The possibility to use the network theory to understand animal collectives has been solved only theoretically, at most to the level of semi-automation~\cite{Gosztolai,Bode,Makagon}. The method proposed in this paper is the first which combines and fully automates 3D individualised object tracking together with networking for potential practical applications.

The approach to collective movement quantification proposed here is informative and easy to interpret. It can be applied to any kind of multiple trajectory data, irrelevant to the species, level or even type of organisation (e.g., for flocks of drones or military ships). The approach has only one adjustable parameter -- the time window selected according to the system type -- making the method model-free, giving a spectrum of networks for all windows.

Even a limited application, such as a single group of a single species, showed a fascinating result: the observed roles in the school are not demonstrated all the time but rather for a time of necessity. This fact addresses a pitfall of many psychological approaches, which assign a single personality type (temperament or MBTI model \cite{IsabelBriggsMyers1995}) to an agent unambiguously and then predict interactions based on this assignment. However, it does not work in this way, and we can see why: even very primitive species show flexible roles dependent on a given situation. It is more suitable to speak about the distribution of such assignments (attributes, categories) or, better, conditional distributions of the roles.

The obtained results also point to pitfalls of classical -- homogeneous -- agent models: even in our simple experiments, the agents received a set of parameters that differed significantly from each other and could be named as \textit{personality} with considerable caution. We believe individual variability is essential to the observed collective phenomena, but we should conduct additional experiments to prove this.

The main advantage of the proposed method, however, is not its generality but its applicability. The method allows converting widespread, well-known tracking data to time-dependent directed graphs (sometimes called coevolution networks \cite{Farajtabar2018,Wang2019}), which are one of the hottest topics of modern data mining \cite{Liu2021} and collective behaviour analysis \cite{Davidson2021a}. Currently, the definition of an agent's role is up to the scientist and his optic. However, the introduction of mathematically defined, graph-based roles can greatly benefit information exchange and convergence between ethological fields and scientific schools. This is a topic for our future research, together with applying this framework to agents at different environmental conditions and levels of biological organisation.
 
The network and hierarchical data extractable by this method can serve as inputs for further modelling and analysis of complex biological systems, including using machine learning or agent-based approaches such as reinforcement learning. The collective analysis described here can be extended to predict the movement of each individual and compare this prediction with the actual behaviour. We expect this method to benefit all parties -- biologists and data scientists -- and encourage further collaboration and knowledge transfer up to the level of commercial and industrial applications, see~\cite{Noldus}. 

\section*{Abbreviations and symbols}
\begin{tabular}{ll}
$A_\textsf{i}$, $A_\textsf{j}$ & position (scalar) of the first object in two consecutive times\\
    & windows \textsf{i} and \textsf{j}, respectively (in our specific case $A \equiv X_i$)\\
$B_\textsf{i}$, $B_\textsf{j}$ & position (scalar) of the second object in two consecutive times\\
    & windows \textsf{i} and \textsf{j}, respectively (in our specific case $B \equiv X_j$)\\
$\vec{A}_\textsf{i}$, $\vec{A}_\textsf{j}$ & vector position of the first agent in frames \textsf{i} and \textsf{j}, respectively (in\\
    & our specific case $\vec{A} \equiv \vec{X}_i$); $\vec{A}_\textsf{i}$ = ($x_{A\textsf{i}}$, $y_{A\textsf{i}}$, $z_{A\textsf{i}}$) and\\
    & $\vec{A}_\textsf{j}$ = ($x_{A\textsf{j}}$, $y_{A\textsf{j}}$, $z_{A\textsf{j}}$) in the 3D Euclidian space\\
$\vec{B}_\textsf{i}$, $\vec{B}_\textsf{j}$ & vector position of the second agent in frames \textsf{i} and \textsf{j}, respectively\\
    & (in our specific case $\vec{B} \equiv \vec{X}_j$); $\vec{B}_\textsf{i}$ = ($x_{B\textsf{i}}$, $y_{B\textsf{i}}$, $z_{B\textsf{i}}$)\\
    & and $\vec{B}_\textsf{j}$ = ($x_{B\textsf{j}}$, $y_{B\textsf{j}}$, $z_{B\textsf{j}}$) in the 3D Euclidian space\\
    $C_{\alpha}$ & $\alpha$-dependent criterion of the Fisher test\\
CNN & convolutional neural network\\      
$D_{\textsf{ij}}$ & difference between the scalar distances in two consecutive time\\
    & frames \textsf{i} and \textsf{j}\\
$e$ & frequency of occurrence of significant evidences in the experiment\\
\textbf{F} & agent's future state (location)\\
\textbf{FF} & follower's and leader's future states correlate (neutral hypothesis)\\
\textbf{FN} & follower's future state correlates with leader's current state\\
    & (supporting hypothesis)\\
\textbf{FP} & follower's future state correlates with leader's past state\\
    & (supporting hypothesis)\\
$i,j$ & labels of agents (fish)\\
\textsf{i},\textsf{j} & labels of two consecutive time windows in the image sequences;\\
    & \textsf{i}, \textsf{j} $\in$ \textsf{N}\\
$i \rightarrow j$ & agent $i$ follows agent $j$\\
$i \not\rightarrow j$ & agent $i$ follows agent $j$\\
$j \rightarrow i$ & agent $j$ follows agent $i$\\
$j \not\rightarrow i$ & agent $j$ follows agent $i$\\
$k$ & number of clusters in $k$-means analysis\\
KDE & kernel density function\\
$L$ & size of the time window\\
\end{tabular}

\begin{tabular}{ll}
$n$ & element of the set $N$\\
$n_{obj}$ & number of agents in the experiment; $n_{obj}$ = 6 fish\\
$N$ & set of combinations of time windows having no relation to the\\
    & hypothesis that one object follows another one;\\
    & $N$ = $\{$\textbf{PP}, \textbf{NN}, \textbf{FF}$\}$\\
\textsf{N} & number of time windows in the sequences\\
\textbf{N} & agent's current state (location)\\
\textbf{NF} & follower's current state correlates with leader's future state\\
    & (rejecting hypothesis)\\
\textbf{NN} & follower's and leader's current states correlate (neutral\\
    & hypothesis)\\
\textsf{N} & number of time windows in the sequences\\
\textbf{N} & agent's current state (location)\\
\textbf{NF} & follower's current state correlates with leader's future state\\
    & (rejecting hypothesis)\\
\textbf{NN} & follower's and leader's current states correlate (neutral\\
    & hypothesis)\\
\textbf{NP} & follower's current state correlates with leader's past state\\
    & (supporting hypothesis)\\
$p$ & $p$-criterion of the Fisher exact test\\
$P$ & follow-up probability that one agent follows another\\
\textbf{P} & agent's past state (location)\\
\textbf{PF} & follower's past state correlates with leader's future state\\
    & (rejecting hypothesis)\\
\textbf{PN} & follower's past state correlates with leader's current state\\
    & (rejecting hypothesis)\\
\textbf{PP} & follower's and leader's past states correlate (neutral\\
    & hypothesis)\\
$r$ & element of the set $R$\\
$R$ & set of combinations of time windows rejecting the hypothesis\\ 
    & that one object follows another one; $R$ = $\{$\textbf{PN}, \textbf{PF}, \textbf{NF}$\}$\\
$s$ & element of the set $S$\\
$S$ & set of combinations of time windows supporting the hypothesis\\
    & that one object follows another one; $S$ = $\{$\textbf{NP}, \textbf{FP}, \textbf{FN}$\}$\\
$t$ & time; the center of the time window\\
VGG & Visual Geometry Group (at Oxford University)\\
$\alpha$ & level of statistical significance\\
\end{tabular}


\vspace{.5cm}

\section*{Supplementary information} The Matlab code and the supplementary video are available at Dryad at~\cite{dryad}. Upon request, we get the image data for further scientific analysis. 

\paragraph{Supplementary 1 -- Video} Time changes of the follow-up networks (i.e., fish school configurations) during the experiment. Top view (\textit{upper left}), bottom view (\textit {upper right}), front view (\textit {lower left}) of fish school in the aquarium, and follow-up network (\textit {lower right}).

\section*{Acknowledgments} This work was supported by the Ministry of Education, Youth and Sports of the Czech Republic--project CENAKVA (LM2018099), from the European Regional Development Fund in the frame of the project ImageHeadstart (ATCZ215) in the Interreg V-A Austria--Czech Republic programme, and by the project GAJU 017/2016/Z.

This version of the article has been accepted for publication, after peer review (if applicable) but is not the Version of Record and does not reflect post/acceptance improvements, or any corrections. The Version of Record is available online at: http://dx.doi.org/10.1007/s40747-023-01057-9.

\section*{Competing interests}
The authors declare that they have no competing interests.

\section*{Author's contributions}
KL collected the image data; he is the algorithm developer and the paper's principal author. RR proofread the paper and contributed critically to the results. D\v{S} is the group leader responsible for funding the research and the inventor of the aquarium device.

\section*{REFERENCES}

\bibliography{Lonhusetal_nature}


\end{document}